\documentclass[a4paper]{article}
\usepackage[lowtilde]{url}
\usepackage{hyperref}
\usepackage{lipsum}
\usepackage{amssymb}

\usepackage[numbers,sort]{natbib} 

\usepackage{INTERSPEECH_v2}

\title{VoxCeleb: a large-scale speaker identification dataset}

\name{Arsha Nagrani$^\dag$, Joon Son Chung$^\dag$, Andrew Zisserman
}
\address{
  Visual Geometry Group, Department of Engineering Science,\\
  University of Oxford, UK}
\email{\{arsha,joon,az\}@robots.ox.ac.uk}

\begin{document}

\maketitle
\begin{abstract}
Most existing datasets for speaker identification contain samples obtained under quite
constrained conditions, and are usually hand-annotated, hence limited
in size. The goal of this paper is to generate a large scale text-independent  speaker identification
dataset collected `in the wild'. 

We make two contributions. First, we propose a fully automated
pipeline based on computer vision techniques to create
the dataset from open-source media.
Our pipeline involves obtaining videos from YouTube; performing active speaker verification using a two-stream synchronization Convolutional Neural Network (CNN), and confirming the identity of the speaker using CNN based facial recognition. We use this pipeline to curate {\tt VoxCeleb} which contains hundreds of thousands of `real world' utterances for over 1,000 celebrities. 

Our second contribution is to apply and compare various state of the art speaker
identification techniques on our dataset to establish baseline
performance. We show that a CNN based architecture obtains the best performance for both identification and verification.

\end{abstract}

\noindent\textbf{Index Terms}: speaker identification, speaker verification, large-scale, dataset, convolutional neural network

\section{Introduction}

\let\thefootnote\relax\footnotetext{\hspace{-12pt}$^\dag$These authors contributed equally to this work.}

Speaker recognition under noisy and unconstrained conditions is an
extremely challenging topic. Applications of speaker recognition are many and varied,
ranging from authentication in high-security systems and forensic
tests, to searching for persons in large corpora of speech data. All
such tasks require high speaker recognition performance under `real
world' conditions. This is an extremely difficult task due to both
extrinsic and intrinsic variations; extrinsic variations include
background chatter and music, laughter, reverberation, channel and
microphone effects; while intrinsic variations are factors inherent to
the speaker themself such as age, accent, emotion, intonation and manner
of speaking, amongst others~\cite{stoll2011finding}. 

Deep Convolutional Neural Networks (CNNs) have given rise to
substantial improvements in speech recognition, computer vision and related fields due to
their ability to deal with real world, noisy datasets without the need
for handcrafted
features~\cite{Krizhevsky12,Simonyan15,he2015deep}. One of the most
important ingredients for the success of such methods, however,  is the
availability of large training datasets.
 
Unfortunately, large-scale public datasets in the field of speaker
identification with unconstrained speech samples are lacking. While
large-scale evaluations are held regularly by the National Institute
of Standards in Technology (NIST), these datasets are not freely
available to the research community. The only freely available dataset curated from
multimedia is the Speakers in the Wild (SITW)
dataset~\cite{mclaren2016speakers}, which contains speech samples of
299 speakers across unconstrained or `wild' conditions. This is a
valuable dataset, but to create it the speech samples have been
hand-annotated. Scaling it further, for example to thousands of
speakers across tens of thousands of utterances, would require the use
of a service such as Amazon Mechanical Turk (AMT). In the computer
vision community AMT like services have 
been used to produce very large-scale datasets, such as ImageNet~\cite{Russakovsky15}.

This paper has two goals. The first is to propose a fully automated
and scalable pipeline for creating a large-scale `real world' speaker
identification dataset. By using visual active speaker identification and
face verification, our method circumvents the need for human
annotation completely. We use this method to curate {\tt VoxCeleb}, a
large-scale dataset with hundreds of utterances for over a thousand
speakers. The
second goal is to investigate different architectures and techniques for training deep CNNs on spectrograms extracted directly from the raw audio files with very little pre-processing, and compare our results on
this new dataset with more traditional state-of-the-art methods. 

{\tt VoxCeleb} can be used for both speaker identification and verification. Speaker identification involves determining which speaker has produced a given utterance, if this is performed for a closed set of speakers then the task is similar to that of multi-class classification. Speaker verification on the other hand involves determining whether there is a match between a given utterance and a target model. We provide baselines for both tasks.

The dataset can be downloaded from \url{http://www.robots.ox.ac.uk/~vgg/data/voxceleb}.

\vspace{-4pt}
\section{Related Works}

For a long time, speaker identification was the domain of Gaussian
Mixture Models (GMMs) trained on low dimensional feature
vectors~\cite{reynolds2000speaker,reynolds1995robust}. The state of the art in more recent times
involves both the use of joint factor analysis (JFA) based
methods which model speaker and channel subspaces separately~\cite{kenny2005joint},
and i-vectors which attempt to model both subspaces into a single
compact, low-dimensional space~\cite{dehak2011front}.
Although state of the art in speaker recognition tasks, these methods all have one thing in common -- they rely on a low dimensional
representation of the audio input, such as Mel Frequency Cepstrum
Coefficients (MFCCs). However, not only does the performance of MFCCs
degrade rapidly in real world noise~\cite{yapanel2002high,hansen2001robust},
but by focusing only on the overall spectral envelope of short
frames, MFCCs may be lacking in speaker-discriminating features (such
as pitch information). This has led to a very recent shift from
handcrafted features to the domain of deep CNNs
which can be applied to higher dimensional inputs~\cite{Sainath15a,hershey2016cnn} and for
speaker identification~\cite{lukic2016speaker}.
Essential to this task however, is a large dataset obtained under real world conditions.

Many existing datasets are obtained under controlled conditions, for
example: forensic data intercepted by police officials~\cite{vannfi}, data from telephone calls~\cite{hennebert2000polycost}, 
speech recorded live in high quality environments such as acoustic
laboratories~\cite{millar1994australian,garofolo1993darpa}, or
speech recorded from mobile devices~\cite{mccool2009mobio,woo2016mit}. \cite{morrison2015forensic}
consists of more natural speech but has been manually processed to
remove extraneous noises and crosstalk.  
All the above datasets are also obtained from single-speaker environments, and are free from audience noise and overlapping speech. 

Datasets obtained from multi-speaker environments include those from
recorded meeting data~\cite{janin2003icsi,mccowan2005ami}, or from audio broadcasts~\cite{bell2015mgb}. These datasets usually contain audio samples under
less controlled conditions. Some datasets contain artificial degradation in an attempt to mimic real world noise, such as those developed using the TIMIT dataset~\cite{garofolo1993darpa}: NTIMIT, (transmitting TIMIT recordings through a telephone handset) and CTIMIT, (passing TIMIT files through cellular telephone circuits). 

Table~\ref{table:existingdata} summarises existing speaker identification datasets.
Besides lacking real world conditions, to the best of our knowledge, most of these datasets have been collected with great manual effort, other than~\cite{bell2015mgb} which was obtained by mapping subtitles and transcripts to broadcast data.

\begin{table}[h!]
\centering
\scriptsize
\begin{tabular}{| l |  r | r | r | r | }
  \hline
~~~~~~~~  \textbf{Name} 							         & \textbf{Cond.}~~~~~~ & \textbf{Free} & \textbf{\# POI}  & \textbf{\# Utter.} \\ \hline 
 ELSDSR~\cite{feng2005new}                       & Clean Speech & \checkmark & 22    &  198    \\ \hline
 MIT Mobile~\cite{woo2016mit}                    & Mobile Devices & - & 88    &  7,884    \\ \hline
 SWB~\cite{godfrey1992switchboard}               & Telephony  & - & 3,114    & 33,039    \\ \hline
 POLYCOST~\cite{hennebert2000polycost}           & Telephony  & - & 133    & 1,285$\ddag$    \\ \hline
 ICSI Meeting Corpus~\cite{janin2003icsi}        & Meetings   & - & 53     & 922    \\ \hline
 Forensic Comparison~\cite{morrison2015forensic} & Telephony  & \checkmark& 552      & 1,264    \\ \hline
 ANDOSL~\cite{millar1994australian}              & Clean speech  & - & 204      & 33,900    \\ \hline
 TIMIT~\cite{fisher1986darpa}$\dag$    & Clean speech  & - & 630 		& 6,300 \\ \hline 

 SITW~\cite{mclaren2016speakers}			 & Multi-media   & \checkmark & 299	& 2,800 \\ \hline 

 NIST SRE ~\cite{greenberg2012nist}  & Clean speech  & -  & 2,000+   & $*$ \\ \hline 

 \textbf{VoxCeleb}							       & Multi-media   & \checkmark & \textbf{1,251}		& \textbf{153,516} \\ \hline 
\end{tabular} 
\vspace{3pt}
\normalsize
\caption{
Comparison of existing speaker identification datasets. 
{\bf Cond.:} Acoustic conditions;
{\bf POI:} Person of Interest; 
{\bf Utter.:} Approximate number of utterances.
\dag And its derivatives.
\ddag Number of telephone calls.
$*$ varies by year.
}
\label{table:existingdata}
\vspace{-18pt}
\end{table}

\vspace{-5pt}
\section{Dataset Description}

{\tt VoxCeleb} contains over 100,000 utterances for 1,251 celebrities, extracted from videos uploaded to YouTube. The dataset is gender balanced, with 55\% of the speakers male. The speakers span a wide range of different ethnicities, accents, professions and ages. The nationality and gender of each speaker (obtained from Wikipedia) is also provided.

Videos included in the dataset are shot in a large number of challenging multi-speaker acoustic environments. These include red carpet, outdoor stadium,  quiet studio interviews, speeches given to large audiences, excerpts from professionally shot multimedia,  and videos shot on hand-held devices. Crucially, all are degraded with real world noise, consisting of background chatter, laughter, overlapping speech, room acoustics, and there is a range 
in the quality of recording equipment and channel noise.  
Unlike the SITW dataset, both audio and video for each speaker is released. 
Table~\ref{table:data_stats} gives the dataset statistics.

\begin{table}[h!]
\centering
\footnotesize
\begin{tabular}{| l | r | }
  \hline
  \textbf{\# of POIs} & 1,251 \\ \hline 
  \textbf{\#  of male POIs} & 690  \\ \hline 
  \textbf{\# of videos per POI} & 36 / 18 / 8 \\ \hline 
  \textbf{\# of utterances per POI} & 250 / 123 / 45 \\ \hline 
  \textbf{Length of utterances (s)} & 145.0 / 8.2 / 4.0 \\ \hline 

\end{tabular} 
\vspace{3pt}
\normalsize
\caption{{\tt VoxCeleb} dataset statistics. Where there are three entries in a field,
numbers refer to the maximum / average / minimum.}
\label{table:data_stats}
\vspace{-18pt}
\end{table}

\vspace{-4pt}
\section{Dataset Collection Pipeline}

\begin{figure*}[ht]
\centering 
\begin{minipage}{.66\linewidth}
\vspace{-5pt}
\fbox{\includegraphics[width=.90\textwidth]{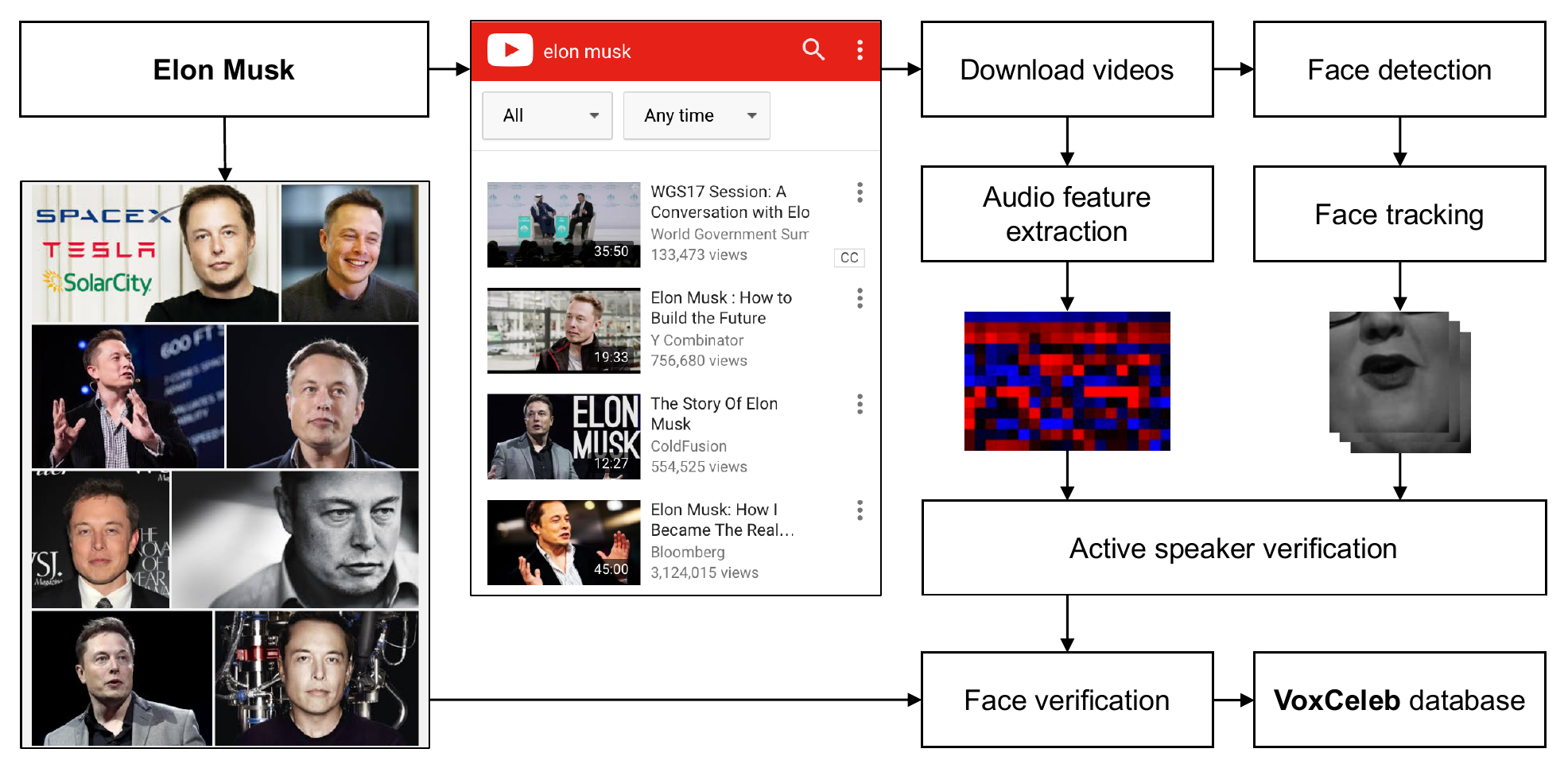} }
\end{minipage}
\begin{minipage}{.33\linewidth}
\includegraphics[width=.99\textwidth]{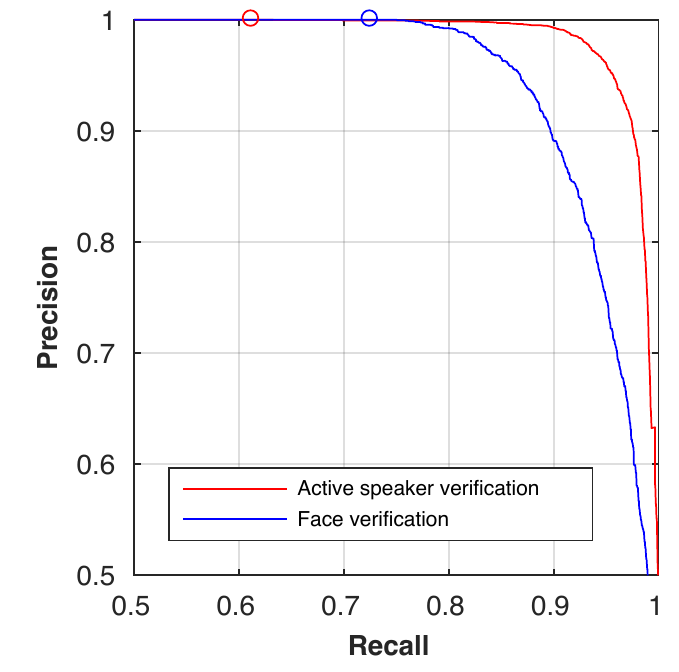}
\end{minipage}
\caption{
{\bf Left:} Data processing pipeline; 
{\bf Right:} Precision-recall curves for the active speaker verification (using a 25-frame window)
and the face verification steps, tested on standard benchmark datasets~\cite{Parkhi15,chakravarty2016cross}. 
Operating points are shown in circles. }
\label{fig:pipeline} 
\vspace{-18pt}
\end{figure*}

This section describes our multi-stage approach for collecting a large speaker recognition dataset, starting from YouTube videos. Using this fully automated pipeline, we have obtained hundreds of utterances for over a thousand different Persons of Interest (POIs). The pipeline is summarised in Figure~\ref{fig:pipeline} left, and key stages are discussed in the following paragraphs: 

\noindent\textbf{Stage 1. Candidate list of POIs.}
The first stage is to obtain a list of POIs. We start from the list of people that appear in the VGG Face dataset~\cite{Parkhi15} , which is based on an intersection of the most searched names in the Freebase knowledge graph, and the Internet Movie Data Base (IMDB). This list contains 2,622 identities, ranging from actors and sportspeople to entrepreneurs, of which approximately half are male and the other half female. 

\noindent\textbf{Stage 2. Downloading videos from YouTube.}
The top 50 videos for each of the 2,622 POIs are automatically downloaded using YouTube search. The word `interview' is appended to the name of the POI in search queries to increase the likelihood that the videos contain an instance of the POI speaking, and to filter out sports or music videos. No other filtering is done at this stage.

\noindent\textbf{Stage 3. Face tracking.}
The HOG-based face detector~\cite{king2009dlib} is used to detect the faces in every frame of the video. Facial landmark positions are detected for each face detection using the regression tree based method of~\cite{kazemi2014one}. The shot boundaries are detected by comparing colour histograms across consecutive frames. Within each detected shot, face detections are grouped together into face tracks using a position-based tracker. This stage is closely related to the tracking pipeline of~\cite{Chung16,Everingham09}, but optimised to reduce run-time given the very large number of videos to process.

\noindent\textbf{Stage 4. Active speaker verification.}
The goal of this stage is to determine the audio-video synchronisation between mouth motion and speech in a video in order to determine which (if any) visible face is the speaker. This is done by using `SyncNet', a two-stream CNN described in~\cite{Chung16a} which estimates the correlation between the audio track and the mouth motion of the video. This method is able to reject the clips that contain dubbing or voice-over. 

\noindent\textbf{Stage 5. Face verification. }
Active speaker face tracks are then classified into whether they are of the POI or not using the VGG
Face CNN. This classification network is based on the VGG-16 CNN~\cite{Simonyan15} trained on the VGG Face
dataset (which is a filtered collection of Google Image Search
results for the POI name). Verification is done by directly using this
classification score with a high threshold.
 
\noindent\textbf{Discussion. }  
In order to ensure that our system is extremely confident that a
person is speaking (Stage 4), and that they have been correctly
identified (Stage 5) without any manual interference, we set very
conservative thresholds in order to minimise the
number of false positives.  Precision-recall curves for both tasks on
their respective benchmark
datasets~\cite{Parkhi15,chakravarty2016cross} are shown in
Figure~\ref{fig:pipeline} right, and the values at the operating point
are given in Table~\ref{table:pr}. Employing these thresholds ensures
that although we discard a lot of the downloaded videos, we can be
reasonably certain that the dataset has few labelling errors.  \\ This
ensures a completely automatic pipeline that can be scaled up to any
number of speakers and utterances (if available) as required.

\begin{table}[h!]
\centering
\footnotesize
\begin{tabular}{| l | l | r | r | }
  \hline
  \textbf{Task}               & \textbf{Dataset} & \textbf{Precision} & \textbf{Recall} \\ \hline 
  Active speaker verification & ~\cite{chakravarty2016cross}    &  1.000    &  0.613 \\ \hline 
  Face verification           & ~\cite{Parkhi15}                &  1.000    &  0.726 \\ \hline 
\end{tabular} 
\normalsize
\caption{Precision-recall values at the chosen operating points.}
\label{table:pr}
\end{table}

\vspace{-4pt}
\section{CNN Design and Architecture}
\label{sec:cnn}

Our aim is to move from techniques that require traditional
hand-crafted features, to a CNN architecture that can choose the features required for the task of
speaker recognition. This allows us to minimise the pre-processing of
the audio data and hence avoid losing valuable information in the
process.

\noindent\textbf{Input features.} 
All audio is first converted to single-channel, 16-bit streams at a
16kHz sampling rate for consistency. Spectrograms are then generated
in a sliding window fashion using a hamming window of width 25ms and step
10ms. This gives spectrograms of size 512 x 300 for
3 seconds of speech.  Mean and variance normalisation
is performed on every frequency bin of the spectrum. This normalisation is
crucial,  leading to an almost 10\%
increase in classification accuracy, as shown in
Table~\ref{table:results_idn}. No other speech-specific preprocessing
(e.g.\ silence removal, voice activity detection, or removal of
unvoiced speech) is used. These short time magnitude spectrograms are
then used as input to the CNN.

\noindent\textbf{Architecture.} 
Since speaker identification under a closed set can be treated as a
multiple-class classification problem, we base our architecture on
the VGG-M~\cite{Chatfield14} CNN, known for good classification performance on image data, with
modifications to adapt to the spectrogram input. 
The fully connected {\em fc6} layer of dimension $9 \times 8$ (support in both dimensions) is replaced by two layers -- a fully connected layer of $9 \times 1$ (support in the frequency domain) and an average pool layer with support $1 \times n$, where $n$ depends on the length of the input speech segment
(for example for a 3 second segment, $n=8$).  
This makes the network invariant
to temporal position but \textit{not} frequency, and at the same time keeps the output dimensions the same as those of the original fully connected layer. This also reduces the number of parameters from 319M in VGG-M to 67M in our network, which helps avoid overfitting. 
The complete CNN architecture is specified in Table~\ref{table:convnet}.

\noindent\textbf{Identification.}
Since identification is treated as a simple classification task, the
output of the last layer is fed into a 1,251-way softmax in order to
produce a distribution over the 1,251 different speakers. 

\noindent\textbf{Verification.}
For
verification, feature vectors can be obtained from the classification
network using the 1024 dimension fc7 vectors, and a cosine
distance can be used to compare  vectors.  
However,  it is better to learn an {\em embedding} by training a
Siamese network with a contrastive loss~\cite{chopra2005learning}. This  is better suited to the verification task as the
network learns to optimize similarity directly,
 rather than indirectly
via a classification loss.
For the embedding network, the last fully connected layer ({\em fc8}) is modified
so that the output size is 1024 instead of the number of classes. We compare both methods in the experiments.

\noindent\textbf{Testing.}
A traditional approach to handling variable length utterances at test time is to 
break them up into fixed length segments (e.g.\
3 seconds) and average the results on each segment to give a final class prediction. 
Average pooling, however allows the network to accommodate variable 
length inputs at test time, 
as the entire
test utterance can be evaluated at once by changing the size of the {\it
apool6} layer.
Not only is this more elegant, it also
leads to an increase in classification accuracy, as shown in
Table~\ref{table:results_idn}.

\begin{table}[ht]
\scriptsize
\centering
\begin{tabular}{| c | c |  c | c | c | c | }
  \hline
  \textbf{Layer}  &  \textbf{Support} & \textbf{Filt dim.} & \textbf{\# filts.} 
  & \textbf{Stride} & \textbf{Data size} \\ \hline 

  conv1 & 7$\times$7 & 1  & 96 & 2$\times$2 & 254$\times$148 \\ \hline
  mpool1 & 3$\times$3 & -  & - & 2$\times$2 & 126$\times$73\\ \hline 
  conv2 & 5$\times$5 & 96  & 256 & 2$\times$2 & 62$\times$36 \\ \hline
  mpool2 & 3$\times$3 & -  & - & 2$\times$2 & 30$\times$17 \\ \hline   
  conv3 & 3$\times$3 & 256  & 384 & 1$\times$1 & 30$\times$17 \\ \hline
  conv4 & 3$\times$3 & 384  & 256 & 1$\times$1 & 30$\times$17 \\ \hline
  conv5 & 3$\times$3 & 256  & 256 & 1$\times$1 & 30$\times$17 \\ \hline
  {\bf mpool5} & 5$\times$3 & -  & - & 3$\times$2 & 9$\times$8 \\ \hline   
  {\bf fc6} & 9$\times$1 & 256  & 4096 & 1$\times$1 & 1$\times$8 \\ \hline
  {\bf apool6} & 1$\times n$ & -  & - & 1$\times$1 & 1$\times$1 \\ \hline
  fc7   & 1$\times$1 & 4096  & 1024 & 1$\times$1 & 1$\times$1 \\ \hline
  fc8   & 1$\times$1 & 1024  & 1251 & 1$\times$1 & 1$\times$1 \\ \hline

\end{tabular} 
\normalsize
\vspace{2pt}
\caption{CNN architecture. 
The data size up to {\it fc6} is for a 3-second input,
but the network is able to accept inputs of variable lengths.
}
\label{table:convnet}
\vspace{-18pt}
\end{table}

\noindent\textbf{Implementation details and training.}
Our implementation is based on the deep learning toolbox MatConvNet~\cite{Vedaldi14}
and trained on a NVIDIA TITAN X GPU. 
The network is trained using batch normalisation~\cite{ioffe2015batch} 
and all hyper-parameters (e.g.\ weight decay, learning rates) use the default values provided with the toolbox.
To reduce overfitting, we augment the data by taking random 3-second crops 
 in the time domain during training. Using a fixed input length is also more efficient.
For verification, the network is first trained for classification
(excluding the test POIs for the verification task, see Section~\ref{sec:exp}),  and then all filter weights are frozen except for the modified last layer and the Siamese network trained with contrastive loss.
Choosing good pairs for training is very important in metric learning.
We randomly select half of the negative examples, and the other half
using Hard Negative Mining, where we only sample from the hardest
10\% of all negatives. 

\vspace{-4pt}
\section{Experiments} 
\label{sec:exp}
{}
This section describes the experimental setup for both speaker identification and verification, and compares the performance of our devised CNN baseline to a number of traditional state of the art methods on {\tt VoxCeleb}. 

\subsection{Experimental setup}

\noindent\textbf{Speaker identification.}
For identification, the training and the testing are performed on the same POIs.
From each POI, we reserve the speech segments from one video for test. The test video contains at least 5 non-overlapping segments of speech. 
For identification, we report \textit{top-1} and \textit{top-5} accuracies.
The statistics are given in Table~\ref{table:tt_idn}.

\noindent\textbf{Speaker verification.}
For verification,  all POIs whose name starts with an `E' are reserved for testing, since this gives a good balance of male and female speakers.
These POIs are not used for training the network, and are only used at test time.
The statistics are given in Table~\ref{table:tt_ver}.

Two key performance metrics are used to evaluate 
system performance for the verification task.
The metrics are similar to those used by existing
datasets and challenges, such as NIST SRE12~\cite{greenberg2012nist}
and SITW~\cite{mclaren2016speakers}. The primary metric is
based on the cost function $C_{det}$ 
\begin{equation}
  C_{det} = C_{miss} \times P_{miss} \times P_{tar} + C_{fa} \times P_{fa} \times (1-P_{tar})
  \label{equ:cdet}
\end{equation}
where we assume a prior target probability $P_{tar}$ of 0.01 and
equal weights of 1.0 between misses $C_{miss}$ and false alarms $C_{fa}$.
The primary metric, $C^{min}_{det}$, is the minimum value of $C_{det}$ for the 
range of thresholds. 
The alternative performance measure used here is 
the Equal Error Rate (EER) which is the rate at 
which both acceptance and rejection errors are equal.
This measure is commonly used for identity 
verification systems.


\begin{table}[h!]
\centering
\footnotesize
\begin{tabular}{| l | r | r | r | }
  \hline
  \textbf{Set} & \# POIs & \# Vid. / POI & \# Utterances \\ \hline 
  \textbf{Dev} & 1,251     &  17.0     &  145,265 \\ \hline 
  \textbf{Test} &  1,251   &   1.0   &  8,251 \\ \hline  \hline 
  \textbf{Total} &  1,251   &   1.0   &  153,516 \\ \hline 

\end{tabular} 
\vspace{3pt}
\normalsize
\caption{Development and test set statistics for identification.}
\label{table:tt_idn}
\vspace{-18pt}
\end{table}

\begin{table}[h!]
\centering
\footnotesize
\begin{tabular}{| l | r | r | r | }
  \hline
  \textbf{Set} & \# POIs & \# Vid. / POI & \# Utterances \\ \hline 
  \textbf{Dev} &   1,211   &   18.0   &  148,642 \\ \hline 
  \textbf{Test} &   40  &    17.4   &  4,874 \\ \hline  \hline 
  \textbf{Total} &   1,251  &    18.0   &  153,516 \\ \hline 

\end{tabular} 
\vspace{3pt}
\normalsize
\caption{Development and test set statistics for verification.}
\label{table:tt_ver}
\vspace{-18pt}
\end{table}

\subsection{Baselines}

\noindent\textbf{GMM-UBM.}
The GMM-UBM system uses MFCCs of dimension 13 as input. Cepstral mean and variance normalisation (CMVN) is applied on 
the features.
Using the conventional GMM-UBM framework, a single speaker-independent universal background model (UBM) of 1024 mixture components is trained for 10 iterations from the training data. 

\noindent\textbf{I-vectors/PLDA.}
Gender independent i-vector extractors~\cite{dehak2011front} are
trained on the {\tt VoxCeleb} dataset to produce 400-dimensional
i-vectors. Probabilistic LDA (PLDA)~\cite{ioffe2006probabilistic} is then used to reduce the dimension of
the i-vectors to 200. 
{}

\noindent\textbf{Inference.} For identification, a one-vs-rest binary SVM classifier is trained for each speaker $m$ (\(m \in 1...K\)).	 All feature inputs to the SVM are L2 normalised and a held out validation set is used to determine the C parameter (determines trade off between maximising the margin and penalising training errors). Classification during test time is done by choosing the speaker corresponding to the highest SVM score. 
The PLDA scoring function~\cite{ioffe2006probabilistic} is used for verification. 


\subsection{Results}

Results are given in Tables~\ref{table:results_idn} 
and~\ref{table:results_ver}. 
For both speaker recognition tasks, the CNN provides superior performance to the traditional 
state-of-the-art baselines. 

For identification we achieve an 80.5\% \textit{top-1} 
classification accuracy over 1,251 different classes, almost 20\% higher than traditional state of the art baselines. 
The CNN architecture uses the average pooling layer for variable length test data.  We also compare to two variants: `CNN-fc-3s', this architecture
has a fully connected fc6 layer,  and divides
the test data into 3s segments and averages the scores. As is evident there is a considerable drop in performance compared to the average pooling original -- partly due to the increased
number of parameters that must be learnt;   `CNN-fc-3s no var.\  norm.', this is the CNN-fc-3s architecture without the variance normalization 
pre-processing of the input (the input is still mean normalized). The difference in performance between the two shows the importance of
variance normalization for this data.

For verification, the margin over the baselines is narrower, but still a significant improvement, with the embedding being the crucial step.

\begin{table}[h!]
\centering
\footnotesize
\begin{tabular}{| l  | r | r | }
  \hline
  \textbf{Accuracy} & Top-1 (\%) & Top-5 (\%) \\ \hline 
  \textbf{I-vectors + SVM}              & 49.0 & 56.6 \\ \hline 
  \textbf{I-vectors + PLDA + SVM}       & 60.8 & 75.6 \\ \hline 
  \textbf{CNN-fc-3s no var.\  norm.}                 & 63.5    & 80.3    \\ \hline 
  \textbf{CNN-fc-3s}         & 72.4   & 87.4    \\ \hline 
  \textbf{CNN}            & {\bf 80.5}    & {\bf 92.1}   \\ \hline 

\end{tabular} 
\vspace{3pt}
\normalsize
\caption{Results for identification on {\tt VoxCeleb} (higher is better).
The different CNN architectures are described in Section~\ref{sec:cnn}.}
\label{table:results_idn}
\vspace{-18pt}
\end{table}

\begin{table}[h!]
\centering
\footnotesize
\begin{tabular}{| l | r | r | }
  \hline
  \textbf{Metrics} & $C^{min}_{det}$ & EER (\%) \\ \hline 
  \textbf{GMM-UBM}                  & 0.80     & 15.0    \\ \hline 
  \textbf{I-vectors + PLDA}       & 0.73     & 8.8    \\ \hline 
  \textbf{CNN-1024D}      & 0.75     & 10.2    \\ \hline 
  \textbf{CNN + Embedding }          & {\bf 0.71}     & {\bf 7.8}    \\ \hline 
\end{tabular} 
\vspace{3pt}
\normalsize
\caption{Results for verification on {\tt VoxCeleb} (lower is better).}
\label{table:results_ver}
\vspace{-18pt}
\end{table}

\vspace{-4pt}
\section{Conclusions}
 We provide a fully automated and scalable pipeline for audio data collection and use it to create a large-scale speaker identification dataset called {\tt VoxCeleb}, with 1,251 speakers and over 100,000 utterances. In order to establish benchmark performance, we develop a novel CNN architecture with the ability to deal with variable length audio inputs, which outperforms traditional state-of-the-art methods for both speaker identification and verification on this dataset. \\


\noindent\textbf{Acknowledgements.} Funding for this research is provided by the EPSRC 
Programme Grant Seebibyte EP/M013774/1 and IARPA grant JANUS. 
We would like to thank Andrew Senior for helpful comments.

\bibliographystyle{ieeetr}


\bibliography{longstrings,vgg_local,vgg_other,mybib}

\end{document}